\documentclass{svmult}
\usepackage{makeidx}         
\usepackage{graphicx}        
\usepackage{multicol}        
\usepackage[bottom]{footmisc}
\makeindex             

\usepackage{amssymb}                       
\usepackage{amsmath}                       
\usepackage{cite}
\newcommand{\RR}{\mathbb{R}}
\newcommand{\CC}{\mathbb{C}}
\newcommand{\ZZ}{\mathbb{Z}}
\newcommand{\NN}{\mathbb{N}}
\newcommand{\semiprod}{>\hspace*{-0.6em}\lhd} 

\newcommand{\Po}{{\mathcal P}_+^{\uparrow}} 
\newcommand{\Poj}{{\mathcal P}_+}
\newcommand{\Lor}{{\mathcal L}_+^{\uparrow}}  
\newcommand{\Lorj}{{\mathcal L}_+}          

\newcommand{\Hlittle}{{\frak h}}                
\newcommand{\calH}{\mathcal{H}}         

\newcommand{\basepoint}{{\bar{p}}}               

\newcommand{\calC}{{\mathcal C}}
\newcommand{\calD}{{\mathcal D}}
          
\newcommand{\calL}{{\mathcal L}}
\newcommand{\calO}{{\mathcal O}}
\newcommand{\calP}{{\mathcal P}}
\newcommand{\calS}{{\mathcal S}}

\newcommand{\Spd}{H}                

\newcommand{\Uirr}{U^{(1)}}                
\newcommand{\Hirr}{\calH^{(1)}}            
\newcommand{\Ulittle}{D}                
\newcommand{\ULor}{V}               
\newcommand{\HLor}{{\frak h}'}               
\newcommand{\LS}{\calD}              
\newcommand{\dom}{{\rm dom }} 
\newcommand{\supp}{{\rm supp}}

\newcommand{\field}{\varphi}   
\newcommand{\Sfield}{\Phi}   
\begin{document}
\title*{String--Localized Covariant Quantum Fields}
\author{Jens Mund } 
\institute{Instituto de F\'{\i}sica, Universidade de 
S\~{a}o Paulo, 
CP 66\,318, 05315 - 970 S\~{a}o Paulo, SP, Brazil
\texttt{mund@fma.if.usp.br}}
\maketitle 
\begin{abstract}
We present a construction of string--localized covariant free quantum
fields for a large class of irreducible (ray) representations of the
Poincar\'e group. Among these are the representations of
mass zero and infinite spin, which are known to be incompatible with
point-like localized fields. (Based on joint work with B.~Schroer and J.~Yngvason~\cite{MSY}.)  
\end{abstract}
\section{Introduction}
\label{secIntro}
The principles of relativistic quantum physics admit certain
``exotic'' particle types  which do not allow for 
point--localized quantum fields, namely 
the massless ``infinite 
spin'' representations found by Wigner~\cite{Wig} and anyons~\cite{Wil}. 
However, it is known~\cite{BGL,M02a} that all Wigner particle types%
\footnote{By Wigner particle type, we mean here an
irreducible unitary ray representation of the Poincar\'e group with
positive energy.} do allow
for localization, in a certain sense, in spacetime 
regions which extend to infinity in some space--like direction. 
 
In this contribution, we present the construction of free 
Wightman type fields for the massless ``infinite spin'' particles, 
which are localized in semi--infinite strings extending to space--like
infinity. This result solves the old problem~\cite{Yng70,IM,Wig48} of
reconciling these representations with the principle
of causality. It has been obtained in collaboration with B.~Schroer and
J.~Yngvason and partly published in \cite{MSY}. The details will be
presented in~\cite{MSY2}. Here, we emphasize  the relation with  
the work of Bros et al.~\cite{BrosMos}, as appropriate for the occasion.  

The construction also works for the usual, ``non--exotic'', particle
types. Our motivation to consider such fields, dispite the fact that
they generate the same algebras as the corresponding point--like
localized free fields, is the hope that they may serve as a starting
point for the construction of interacting string--localized quantum
fields. 

Let us make precise what we mean by a string--localized covariant free
quantum field for a given particle type.%
\footnote{Our notion of a  string--localized covariant quantum field is a generalization 
of the generalized Wightman fields of Steinmann~\cite{St}.} 
The ``{\em string}'' is a ray which extends from a point $x\in\RR^d$ to 
infinity in a space--like direction. That is to say, it is 
of the form  $x+\RR^+e$, where $e$ is in the manifold of space--like directions 
\begin{equation} \label{eqSpd} 
\Spd^{d-1} :=\{ e\in\RR^d:\, e\cdot e=-1\}.  
\end{equation}
Let now $U$ be a unitary ray representation of the Poincar\'e group acting on a Hilbert 
space $\calH$ with positive energy and a unique invariant vector $\Omega$, 
which contains an irreducible  ray representation $\Uirr$ acting on $\Hirr\subset \calH$. 
%
\begin{definition} \label{StringLocField} 
A {\em string--localized covariant quantum field} for $\Uirr$ is an operator  
valued distribution $\field(x,e)$ over $\RR^d\times \Spd^{d-1}$ 
acting on $\calH$    
such that the following requirements are satisfied. 
\\[1ex]
0) {\em Reeh--Schlieder property:} 
$\Omega$ is cyclic for the polynomial algebra of fields $\field(f,h)$ with 
$\supp f\times\supp h$ in a fixed region in $\RR^d\times\Spd^{d-1}$. 
\\[1ex]
i) {\em Covariance:} For all $(a,\Lambda)\in\Po$ and $(x,e)\in \RR^d\times
\Spd^{d-1}$ holds 
\begin{equation} \label{eqFieldCov}
U(a,\Lambda)\field(x,e)\,U(a,\Lambda)^{-1}=\field(\Lambda x+a,\Lambda e).
\end{equation}
ii) {\em String--locality:} If the strings 
$x_1+\RR^{+}e'_1$  and $x_2+\RR^{+} e_2$ are
space--like separated  for all $e'_1$ in some open neighborhood of
$e_1$, then 
\begin{equation} \label{eqFieldLoc}
[\field(x_1,e_1),\field(x_2,e_2)]=0. 
\end{equation}
The field is called {\em free}, if it creates only single particle states
from the vacuum vector, $ \field(f,h)\Omega \in \Hirr$.
\end{definition}

Our construction of such fields in \cite{MSY,MSY2} is reduced to a 
single particle problem. Namely, consider the single particle vector
$\psi(x,e):=\field(x,e)\Omega$ if a free field $\field(x,e)$ as above is given. It 
enjoys certain specific properties reflecting 
the covariance and locality of the field. 
The crucial point is that these properties are intrinsic to the representation 
$\Uirr$ and can be formulated 
without reference to the field, using the 
concept of a modular localization structure~\cite{BGL,FasS02,M02a}
based on Tomita--Takesaki modular theory.  We will call a $\Hirr$-valued distribution 
satisfying the ensuing properties a {\em string--localized covariant wave
function} for $\Uirr$, cf.\ Definition~\ref{StringLocWFct}. 
Our strategy is to reverse the route, namely to construct such a
$\Hirr$-valued distribution $\psi(x,e)$
for given $\Uirr$ and then to obtain the field via second quantization. 

The idea of the construction of $\psi(x,e)$ is as follows. 
Recall that an irreducible representation $\Uirr$ of the Poincar\'e group is
induded by a representation $\Ulittle$ of a subgroup $G$ of the
Lorentz group. 
If $\ULor$ is an extension of $\Ulittle$ to the Lorentz group, then $\Uirr$ is contained in
$U_0\otimes\ULor$, where $U_0$ is the scalar representation. Thus the
problem can be separated. The $U_0$ part is solved by Fourier
transformation. 
Now Bros et al. exhibit in~\cite{BrosMos} a suitable  
representation $\ULor$, for which they
(implicitely) construct a localized covariant wave function living on $\Spd^{d-1}$. 
Consider then the tensor product of a wave 
function localized at $x$ for $U_0$ and a wave function localized
at $e\in\Spd^{d-1}$ for $\ULor$. Our basic result is that the projection
onto $\Uirr$ of this vector turns out to be a vector which is
localized for $\Uirr$ in the  string
with initial point $ x$ and direction $e$. 

We recall the relevant representations $\Uirr$ of the Poincar\'e group and 
the concept of a modular localization structure in Sections 2 and 3,
respectively. We will concentrate on the bosonic
representations with positive mass and for those with
zero mass and infinite spin, in dimension $d=3$ and $4$. 
In Section 4 we present the (definition and) construction
of a string--localized covariant wave function, as sketched above. 
In Section 5 we summarize our results and give a brief outlook. 
\section{Wigner Particles} \label{secWigPart}
Following Wigner~\cite{Wig}, the state space of an elementary
relativistic particle  corresponds to an irreducible ray representation of the  
Poincar\'e group with positive energy. 
We recall the relevant representations here for spacetime
dimension $d=3$ and $4$, restricting to proper representations since we
are at the moment only interested in bosons. 
We denote the proper orthochronous Poincar\'e and Lorentz groups by
$\Po$ and $\Lor$, respectively. Reflecting the semidirect product
structure $\Po=\RR^d\semiprod\Lor$, elements of the Poincar\'e group will
be denoted $g=(a,\Lambda)$. 

An irreducible positive energy representation $\Uirr$ of $\Po$ is characterized by 
two data. The first one is the mass value 
$ m\geq0$, determining the energy--momentum spectrum of the
corresponding particle as the mass hyperboloid 
\begin{equation} \label{eqMassHyp}
H_m^+ :=\{p\in\RR^d:\, p\cdot p=m^2, \, p_0>0\}.  
\end{equation}
Given $m$, one fixes a base point $\basepoint\in H_m^+$, and considers
the stabilizer subgroup, within $\Lor$, of this point. This so--called
``little group'' will be denoted $G_\basepoint$ in the sequel. Then
the second characteristicum of $\Uirr$ is a unitary irreducible
representation $V$ of $G_\basepoint$, acting in a Hilbert space $\Hlittle$.  
The representation $\Uirr$ fixed by these data is said to be {induced} from
$\Ulittle$. It acts on 
\begin{equation} \label{eqHS}
\Hirr:= L^2(H_m^+,\D\mu)\otimes \Hlittle,
\end{equation} 
which we identify with $L^2(H_m^+,\D\mu;\Hlittle)$, according to 
\begin{eqnarray} \label{eqUirr} 
\big(\Uirr(a,\Lambda)(\phi\otimes\varphi)\big)(p)&= e^{ \I a\cdot
p} \,\phi(\Lambda^{-1}p)\, \Ulittle({R(\Lambda,p)})\varphi. 
\end{eqnarray}
Here, $R(\Lambda,p)$ is the so--called Wigner rotation, defined by 
\begin{equation}  \label{eqWigRot} 
R(\Lambda,p):= A_p^{-1}\,\Lambda\,A_{\Lambda^{-1}p}, 
\end{equation}
where $(A_p,\,p\in H_m^+)$ is a section of the bundle $\Lor\rightarrow
H_m^+$, i.e.\ $A_p$ maps $\basepoint$ to $p$.  

The little groups $G_\basepoint$ can be conveniently determined as follows. Let 
\begin{align} \label{eqOrbit}
\Gamma_\basepoint :=\{ q\in H_0^+:\, q\cdot \bar p=1\}.  
\end{align}
Then $G_\basepoint$ is precisely the (unit component of the) isometry group
of $\Gamma_\basepoint$. But $\Gamma_\basepoint$, with the induced metric from ambient
Minkowski space, is  isometric to the sphere
$S^{d-2}$ for $m>0$,  and to $\RR^{d-2}$ for $m=0$. (Eg.\ for $m=0$ and $d=4$, the map 
$\xi:\RR^{2}\rightarrow \Gamma_\basepoint$, with $\basepoint =
\frac{1}{2}(1,1,0,0)$, defined by 
\begin{equation} \label{eqGammaR}
\xi(z):= (|z|^2+1,|z|^2-1,z_1,z_2\big) 
\end{equation}
can be checked to be an isometric diffeomorphism.) It follows that the little group
$G_\basepoint$ is for $m>0$ isomorphic
to $SO(d-1)$,  
and for $m=0$ isomorphic to the euclidean group in $d-2$
dimensions, i.e.\ $G_\basepoint\cong E(2)$ in $d=4$ and
$G_\basepoint\cong \RR$ in $d=3$.  Now {faithful} representations of  $E(2)$ are infinite
dimensional. Owing to this fact, a representation $\Uirr$ resulting from
$m=0$ and a faithful representation of $G_\basepoint$ is called a {\em massless infinite
spin representation}. 
The faithful representations $\Ulittle$ of $E(2)$ are labelled by a 
strictly positive number $\kappa>0$,
and $\Ulittle=\Ulittle_{(\kappa)}$ acts on $L^2(\RR^2,\delta(|k|^2-\kappa^2))$
according to 
\begin{equation} \label{eqUlittle}
\big(\Ulittle_{(\kappa)}(c,R)\tilde{\varphi}\big)(k) := 
\exp({\I c\cdot k})\,\tilde{\varphi}(R^{-1}k),\quad (c,R)\in E(2). 
\end{equation}
%
\section{Modular Localization Structure for ${\Uirr}$} 
As the first step in our construction, the single particle space is endowed with a family of
so--called Tomita operators, 
labelled by a specific class of spacetime regions. This family will be
called a modular localization structure for the single particle space. 

The basic geometrical ingredient is the family of wedge regions. 
A {\em wedge} is a region in Minkowski space which arises by a
Poincar\'e transformation from the ``standard wedge'' 
$$
W_0:=\{ x\in \RR^d: |x^0|<x^{1}\}. 
$$ 
Associated with each wedge $W$ is the one--parameter group of Lorentz boosts $\Lambda_W(t)$
leaving $W$ invariant, and the reflection $j_W$ about the edge of $W$. 
More precisely, for the standard wedge $W_0$ the boosts 
$\Lambda_{W_0}(t)$ act on the coordinates $x^0,x^{1}$ as 
\begin{equation} \label{eqBoox}  
\left( \begin{array}{cc}
 \cosh(t) &  \sinh(t)  \\
 \sinh(t) &  \cosh(t)
 \end{array} \right), 
\end{equation} 
and the reflection $j_{W_0}$ inverts the sign of the coordinates
$x^0,x^{1}$ and leaves the other coordinate(s) invariant. 
For a general wedge $W=g\,W_0$, $g\in\Po$, the boosts and reflection
are defined as\footnote{This definition is consistent 
because every Poincar\'e transformation which leaves $W_0$ invariant
commutes with $\Lambda_{W_0}(t)$ and $j_{W_0}$, cf.~\cite{BGL}.}  
\begin{eqnarray} \label{eqLorW}
\Lambda_{gW_0}(t):=& g\,\Lambda_{W_0}(t)\,g^{-1},\\
j_{gW_0}         :=& g\, j_{W_0}\,g^{-1}. \label{eqJW}
\end{eqnarray}

Let now $U$ be an (anti-) unitary representation of the proper
Poincar\'e group acting in some Hilbert space $\calH$. 
Then there is, in particular, for each wedge $W$ an anti-unitary representer $U(j_W)$ of the
reflection $j_W$.  Let further $K_W$ denote the self-adjoint generator
of the unitary group representing the corresponding boosts, i.e.\
$K_W$ is defined by $\exp(\I tK_W)=U(\Lambda_W(t))$ for all
$t\in\RR$. Then we define an anti--linear  operator associated with $W$ by  
\begin{equation} \label{eqSW} 
S_U(W):= \;U(j_W) \exp({-\pi K_W}). 
\end{equation}
Owing to the group relations, it is an antilinear involution,
$S_U(W)^2\subset 1$, i.e.\ a so--called Tomita operator. 

We now consider the class of causally complete, convex spacetime
regions, which we denote by $\calC$. It is known~\cite{BGL} that each
$C\in\calC$ coincides with the intersection of all wedges which
contain $C$. Typical regions belonging to this class are double cones,
space--like cones, and wedges. 
For each $C\in\calC $ we now define the subspace of vectors which are ``localized in $C$'' by 
\begin{equation}  \label{eqLS}
 \LS_U(C):= \big\{ \psi \in \bigcap_{W \supset C} \dom S_U(W), \;
  S_U(W)\psi \mbox{ independent of } W\big\}. 
\end{equation}
Brunetti et al.\ have shown~\cite{BGL} that if $U$ has positive
energy, then $\LS_U(C)$ is dense in $\calH$ if $C$ contains a space--like cone.  
On this domain we define a closed anti--linear involution
$S_U(C)$ by\footnote{We shall skip the index $U$ when no confusion can
arise.} 
\begin{equation}  \label{eqSC}
 S_U(C)\psi := S_U(W)\psi, \quad W\supset C. 
\end{equation}
The family of these anti--linear involutions satisfies isotony~\cite{BGL}, 
$S_U(C_1)\subset S_U(C_2)$ for  $C_1\subset C_2$, and covariance, 
$U(g)S_U(C)U(g)^{-1}=S_U(gC)$. It has further a property~\cite{BGL}
which will soon turn out to correspond to locality: 
\begin{lemma} \label{TomOp} 
If $C_1$ and $C_2$ are causally disjoint, then 
\begin{equation} \label{eqSS*}
S_U(C_1)\subset S_U(C_2)^*.
\end{equation}
\end{lemma}
\begin{proof}
Choose a wedge $W$ which contains $C_1$ and
whose  causal complement $W'$ contains $C_2$. 
The group relations  $\Lambda_{W'}(t)=\Lambda_{W}(-t)$
and $j_{W'}=j_W$, cf.~\cite{BGL}, imply that $S({W'})=U(j_W)\exp(\pi K_W)$. On the other
hand, $\Lambda_{W}(t)$ commutes with $j_W$, hence 
$U(j_W)\exp(\pi K_W)$$=\exp(-\pi K_W)U(j_W)$$\equiv S(W)^*$. Hence 
$S(W')=S(W)^*$. Therefore 
$$
S(C_1)\subset S(W)=S(W')^*\subset S(C_2)^*,
$$
which proves the claim. 
\smartqed \qed
\end{proof}
All these properties motivate us to call the family $S_U(C)$,
$C\in\calC$, a {\em modular localization structure}%
\footnote{Note that our notion of a modular localization
structure is equivalent to the usual one, as formulated e.g.\ in 
\cite{BGL,FasS02,M02a,MSY}. There, one considers for each $C\in\calC$ the real subspace 
\begin{equation*} 
K_U(C):= \bigcap_{W\supset C} \{\phi\in\dom S_U(W):\; S_U(W)\phi=\phi\}. 
\end{equation*}
These are precisely the $+1$ eigenspaces of our Tomita
operators~\eqref{eqLS}, \eqref{eqSC}. But these eigenspaces are well--known~\cite{LRT,RvD} 
to be in one-to-one correspondence with the latter. }
for the representation $U$.  They allow the construction of a local
and covariant theory for a given particle type from the
single particle space via second quantization as follows.  
Given the corresponding irreducible representation $\Uirr$ of $\Po$, 
extend it to $\Poj$ as eg.\ in Appendix~\ref{Poj},  
and define $\LS(C)=\LS_{\Uirr}(C)$ as above. 
Let $a^*(\psi)$ and $a(\psi)$, for $\psi\in\Hirr$, denote the creation
and annihilation operators acting on the symmetrized Fock space over
$\Hirr$. 
Then define, for $\psi\in\LS(C)$, 
\begin{equation}  \label{eqSfield}
\Sfield(\psi):= a^*(\psi) + a(S(C)\psi) .
\end{equation} 
These operators generate a covariant and local theory~\cite{BGL}, the locality
property coming about as follows. 
For $\psi_1\in\LS(C_1)$, $\psi_2\in\LS(C_2)$, the commutator 
$[\Sfield(\psi_1),\Sfield(\psi_2)]$ equals 
$ (S(C_1)\psi_1,\psi_2) - (S(C_2)\psi_2, \psi_1)$. But if $C_1$ is
causally disjoint from $C_2$, this expression vanishes by Lemma~\ref{TomOp}, hence 
\begin{equation}\label{eqSfieldLoc}
[\Sfield(\psi_1),\Sfield(\psi_2)]=0   
\end{equation}
in this case. Thus, the property \eqref{eqSS*} of our modular localization
structure implies the locality property of the second quantization. 

The motivation to construct the modular localization structure from
the representation $U$ is the Bisognano-Wichmann
theorem~\cite{BiWi,M01a}. This theorem states that for a large class of local
relativistic quantum fields $\field(f)$ the so-called modular
covariance property holds:  
$$ 
 S(C)\field(f)\Omega= \field(f)^*\Omega \;\mbox{ if } C\supset\supp f,
$$ 
where $S(C)$ is constructed as above, cf.\ \eqref{eqSW} to
\eqref{eqSC}, from the representation $U$ under
which the  field is covariant. Thus, 
given a local quantum field  $\field(f)$, the vectors $\field(f)\Omega$, $\supp
f\subset C$, are the prototypes for elements of the subspace $\LS_U(C)$. 

\section{String--Localized Covariant Wave Functions}
In view of the above discussion, our task of constructing a 
string--localized covariant free quantum field for a given particle
type reduces to the first-quantized version of the problem:  Namely, the
construction of the spaces $\LS(C)$ in terms of
``covariant string--localized wave functions'' as mentioned in the
introduction. These are defined as follows. Let 
$\Uirr$ be the corresponding representation, acting on $\Hirr$.  
\begin{definition} \label{StringLocWFct} 
A {\em string--localized covariant wave function} for $\Uirr$ is a 
weak $\Hirr$-valued distribution $\psi(x,e)$ on $\RR^d\times \Spd^{d-1}$ 
satisfying the following requirements. 
\\[1ex]
0) The set of $\psi(f,h)$, with $\supp f\times\supp h$ in a fixed
compact region in $\RR^d\times\Spd^{d-1}$, is dense in $\Hirr$. 
\\[1ex]
i) {\em Covariance:} For all $(a,\Lambda)\in\Po$ and $(x,e)\in \RR^d\times
\Spd^{d-1}$ holds  
\begin{equation} \label{eqWFCov}
\Uirr(a,\Lambda)\psi(x,e)=\psi(\Lambda x+a,\Lambda e).
\end{equation}
ii) {\em String--locality:} 
If $\supp f + \RR^+ \supp h\subset C\in\calC$, then 
$\psi(f,h)$ is in $\LS_{\Uirr}(C)$. 
\end{definition} 
Given such $\psi(x,e)$, one verifies that  
\begin{equation} \label{eqFieldSfield}
\field(x,e):= \Sfield(\psi(x,e)) ,
\end{equation}
with $\Sfield(\psi)$ as in \eqref{eqSfield}, is a string--localized
covariant free quantum field in the sense of Definition~\ref{StringLocField}. 
(Locality~\eqref{eqFieldLoc} follows from~\eqref{eqSfieldLoc}.)  
\begin{example} \label{LSU0} 
To illustrate the concept, we consider the scalar irreducible  
unitary representation  $U_0$ with mass $m\geq0$. (Scalar means that the little
group is represented trivially.) For $f\in\calS(\RR^d)$, let 
$Ff$ denote the restriction of the Fourier transform of
$f$ to the mass shell $H_m^+$.  
This map enjoys the covariance properties 
\begin{eqnarray} \label{eqFCov} 
U_0(g) \,Ff &=& F\,{g_*f}, \quad g\in\Po,  \\
U_0(j) \,Ff &=& F\,j_*\bar{f}, \quad j\in\calP_+^{\downarrow}, \label{eqFCovj}
\end{eqnarray}
where $(g_*f)(x):=f(g^{-1}x)$. Further, 
if $f$ has compact support contained  in some wedge $W$, then 
\begin{equation} 
\exp(-\pi (K_0)_W) Ff= F(j_W)_*f, \label{eqFModOp} 
\end{equation}
where $(K_0)_W$ is the generator of $U_0(\Lambda_W(t))$. 
The basic fact underlying this identity is that for $x\in W$, the analytic 
function $t\mapsto\Lambda_W(t) x$ has imaginary part in the forward light cone for
$t$ in the strip $\RR+i(0,\pi)$, and goes to $j_W x$ if $t$ goes to
$i\pi$. Lemma~\ref{ModOp} in the appendix then implies~\eqref{eqFModOp}.  
It follows that $S_{U_0}(W) Ff=F\bar{f}$, hence $Ff\in\LS_{U_0}(\calO)$
if $\supp f\subset\calO$. 
Consequently the map $f\mapsto Ff$ is, in analogy to the above
definition,  a (point-) localized covariant wave function for $U_0$. 
Note that the definition \eqref{eqFieldSfield}, namely $\field(f):=\Sfield(Ff)$ 
with  $\Sfield(\cdot)$ as in \eqref{eqSfield}, then coincides 
with the usual scalar free field.  
\end{example}
We now turn to the construction of a {string}--localized covariant
wave  function for arbitrary $\Uirr$ with mass $m\geq0$ and faithful
(or scalar) inducing representation $\Ulittle$ of the little group $G_\basepoint$. 
Bros et al.\ exhibit in \cite{BrosMos} a family of unitary irreducible
representations $\ULor^\alpha$ of the Lorentz group $\Lorj$, labelled
by a complex number $\alpha$ with real part $-(d-2)/2$.\footnote{$\ULor^\alpha$
is in fact equivalent to the irreducible principal series
representation corresponding to the value $-|\alpha|^2$ of the
Casimir operator.}
As we show in Lemma~\ref{UlittleInULor}, the inducing representation $\Ulittle$ is contained in
the restriction of $\ULor^\alpha$ to $G_\basepoint$, namely as a subrepresentation if
$m>0$ and in a direct integral decomposition if $m=0$. 
This implies that $\Uirr$ is contained in the representation 
induced by $\ULor^\alpha|G_\basepoint$. But the latter is equivalent to 
the representation 
$U_0 \otimes\ULor^\alpha$, hence $\Uirr$ is contained in $U_0
\otimes\ULor^\alpha$. More precisely, there is a map $R^\alpha$ from (a dense
domain in) the tensor product of the representation spaces of $U_0$
and $\ULor^\alpha$ into the representation space of $\Uirr$ 
satisfying 
the intertwiner relation 
\begin{equation} \label{eqIntUirrUtens} 
\Uirr(a,\Lambda) \circ R^\alpha =
R^\alpha \circ \;U_0(a,\Lambda)\otimes \ULor^\alpha(\Lambda) ,\quad
(a,\Lambda)\in\Po, 
\end{equation} 
on its domain. We write down a suitable intertwiner $R^\alpha$ in 
Lemma~\ref{UlittleInULor}, which turns out to satisfy also 
\begin{equation} \label{eqIntUirrUtensJ} 
\Uirr(j) \circ R^\alpha =
R^{\bar{\alpha}} \circ \;U_0(j)\otimes \ULor^\alpha(j) ,\quad
j\in\calL_+^{\downarrow}. 
\end{equation} 
Thus, the  problem of finding a string--localized wave function can now be
separated.  For $U_0$ we already have a localized wave function, cf.\
Example~\ref{LSU0}. Now for $\ULor^\alpha$, Bros et al.~\cite{BrosMos} 
construct implicitely a ``localized covariant wave function'' on
$\Spd^{d-1}$, in the following sense: 
\begin{example} \label{LSULor}
There is a continuous linear map $F^\alpha$ from $\calD(\Spd^{d-1})$ into 
the representation space of $\ULor^\alpha$ with the following properties: 
\\
{\em 0)} The set of $F^\alpha h$, with $\supp h$ in a fixed region in
$\Spd^{d-1}$, is dense. 
\\
{\em i)} For  $h\in \calD(\Spd^{d-1})$ holds 
\begin{eqnarray}  \label{eqFHCov} 
\ULor^\alpha(\Lambda)\,F^\alpha h &=& F^\alpha \Lambda_*h,\quad  \Lambda\in\Lor,\\
\ULor^\alpha(j)\,F^\alpha h &=& F^{\bar{\alpha}} j_*\bar{h},\quad  j\in\calL_+^{\downarrow}.
 \label{eqFHCovj} 
\end{eqnarray}
{\em ii)} For a wedge $W$ whose edge contains the origin, let $K^\alpha_{{W}}$ be the
generator of $\ULor^\alpha(\Lambda_W(t))$. Then for all $h\in\calD(\Spd^{d-1})$
with $\supp h\subset W\cap \Spd^{d-1}$, the vector $F^\alpha h$ is in the
domain of $\exp(-\pi K^\alpha_{{W}})$, and 
\begin{equation} \label{eqFHModOp}
 \exp(-\pi K^\alpha_{{W}})\;F^\alpha h = F^\alpha \,(j_{W})_* h.
\end{equation}
(We recall the definition of $F^\alpha$ in the appendix, cf.~\eqref{eqFH}, 
and show in Lemma~\ref{FHProp'} that the mentioned properties are implicitly 
contained in~\cite{BrosMos}.) 
\end{example}
All this implies that a good candidate for a covariant
string--localized wave function in the sense of
Definition~\ref{StringLocWFct} is given by 
\begin{equation} \label{eqWF}
\psi^\alpha(f,h) := R^\alpha (F f\otimes F^\alpha h), 
\end{equation} 
with $f\in\calD(\RR^{d})$ and $ h \in\calD(\Spd^{d-1})$. 
We have in fact the following result. 
\begin{proposition}  \label{WF} 
Equation~\eqref{eqWF} defines a string--localized covariant wave
function for $\Uirr$ in the sense of  Definition~\ref{StringLocWFct}. 
Moreover, for $C\in\calC$ the Tomita operator  $S(C)$ acts as follows. 
Let $\calO\subset\RR^d$ and $\hat{\calO}\subset \Spd^{d-1}$ be
such that $\calO+\RR^+\hat{\calO}\subset C$. Then for all $f$ with 
$\supp f\subset \calO$ and $h$ with $\supp h\subset \hat{\calO}$, 
\begin{equation} \label{eqSPsifh}
S(C) \, \psi^\alpha(f,h)  =\psi^{\bar\alpha}(\bar f,\bar h) .
\end{equation}
For $m=0$, $\psi^\alpha(f,h)$ has the explicit form 
\begin{equation}  \label{eqWigCovfh} 
\psi^\alpha(f,h)(p) = Ff(p)\,u^\alpha(h,p) ,
\end{equation}
where $h\mapsto u^\alpha(h,p)$ is the $\Hlittle$-valued distribution  
on $\Spd^{d-1}$ with kernel 
\begin{equation} \label{eqIntDist}
u^\alpha(e,p)(k)= \E^{-\I\pi\alpha/2}\, \int_{\RR^{d-2}}\D z e^{\I k  z}\, 
\big(\xi(z)\cdot A_p^{-1}e\big)^\alpha. 
\end{equation}
Here $z\mapsto \xi(z)$ is the isometry from $\RR^{d-2}$ onto
$\Gamma_\basepoint$ exhibited in \eqref{eqGammaR}. 
\end{proposition}
\begin{proof} 
We first consider $m>0$,  in which case the intertwiner $R^\alpha$ is a partial
isometry defined on the whole Hilbert space, cf.\ Lemma~\ref{UlittleInULor}. 
Then  the covariance condition $i)$ of  Definition~\ref{StringLocWFct}
is satisfied  by construction,  cf.\ \eqref{eqFCov},
\eqref{eqIntUirrUtens} and \eqref{eqFHCov}. 
The ``Reeh-Schlieder'' property {\em 0)} of Definition~\ref{StringLocWFct}
follows from the well-known  Reeh--Schlieder property of  $Ff$ and
that of $F^\alpha h$, cf.\ {\em 0)} of Example~\ref{LSULor}. 
It remains to prove \eqref{eqSPsifh}, which then implies the locality
property $ii)$. 
As a first step, let $f$ and $h$ be such that $\supp f+\RR^+\supp h$
is contained in the standard wedge $W_0$. 
It then follows that $\supp f$ is contained in $W_0$ and $\supp h$ in
its {\em closure}. 
Suppose first that $\supp h\subset W_0$. 
Then from \eqref{eqFModOp} and \eqref{eqFHModOp} we know that the
vectors $Ff$ and $F^\alpha h$ are in the
domains of the corresponding ``modular operators'' $\exp(-\pi K_{W_0})$ and
that the latter maps them to $F(j_0)_*f$ and $F^\alpha(j_0)_*h$,
respectively. From the intertwining property \eqref{eqIntUirrUtens} of $R^\alpha$
and its  continuity it follows (eg.\ using Lemma~\ref{ModOp}) that
$\psi^\alpha(f,h)$ is in the domain of $\exp(-\pi K_{W_0})$ and that   
\begin{equation} \label{eqModOpPsifh}
\exp(-\pi K_{W_0})\;\psi^\alpha(f,h) =\psi^\alpha\big((j_0)_*f,(j_0)_*h\big). 
\end{equation}
Further, \eqref{eqFCovj}, \eqref{eqIntUirrUtensJ} and \eqref{eqFHCovj}
imply that 
$\Uirr(j)\,\psi^\alpha(f,h) =
\psi^{\bar{\alpha}}\big(j_*\bar{f},j_*\bar{h}\big)$ for $j\in\calL_+^\downarrow$.  
Now the last two equations imply that 
\begin{equation} \label{eqSPsifh'}
S(W_0) \, \psi^\alpha(f,h)  =\psi^{\bar\alpha}(\bar f,\bar h).
\end{equation}
If, on the other hand, $\supp h$ meets the boundary of $W_0$ (but is
contained in its closure), then one finds a sequence $h_n\to h$ so that
$\supp h_n\subset W_0$ for all $n$. Then $S(W_0) \psi^\alpha(f,h_n)$
goes to $\psi^{\bar\alpha}(\bar f,\bar h)$, and \eqref{eqSPsifh'} also
holds in this case because $S(W_0)$ is closed. By covariance, it
follows that for any wedge $W$, the operator $S(W)$ acts as in
\eqref{eqSPsifh'} if $\supp f+\RR^+\supp h\subset W$. This proves
\eqref{eqSPsifh}. The continuity property follows from that of $F$, $F^\alpha$ and 
$R^\alpha$. The proof is complete for $m>0$. 

For $m=0$, we show in~\cite{MSY2} the following facts. 
$R^\alpha$ is well-defined on vectors of the form 
$Ff\otimes F^\alpha h$, leading to the formula~\eqref{eqWigCovfh}, \eqref{eqIntDist}, 
and the intertwining properties \eqref{eqIntUirrUtens} and 
\eqref{eqIntUirrUtensJ} hold on these vectors.  
Further, if $h$ has support in a wedge $W$, then for almost all $p$ the
$\Hlittle$-valued function $t\mapsto u^\alpha(\Lambda_W(t)_*h,p)$ is analytic on
the strip $\RR+\I(0,\pi)$ and weakly continuous on its closure. It is 
uniformly bounded in $p$ and $t$, for $p$ in a dense set of 
$H_0^+$ and for $t$ in any compact subset of the closure of the strip. 
As $t$ goes to $\I\pi$, it goes to $u((j_W)_*h,p)$. 
This implies \eqref{eqModOpPsifh}, eg.\ using Lemma~\ref{ModOp}
(details are spelled out in~\cite{MSY2}). The proof of \eqref{eqSPsifh} is then completed as
in the case $m>0$. Finally, we show in~\cite{MSY2} an analyticity and growth property 
in $e$ of $u^\alpha(e,p)$ 
which implies continuity of $\psi^\alpha(f,h)$. 
\smartqed\qed
\end{proof}
\section{Summary and Outlook} 
We have constructed, for each $\alpha\in\CC$ with $\Re \alpha=-(d-2)/2$ and 
each massless ``infinite spin'' representation $\Uirr$, a $\Hirr$-valued distribution 
on $\RR^d\times \Spd^{d-1}$ with certain specific 
properties, which motivate our name ``string--localized covariant wave
function''. They guarantee that second quantization~\eqref{eqFieldSfield} of these object
leads to a string--localized covariant free quantum field, 
cf.~Definition~\ref{StringLocWFct} and discussion thereafter. 
Summarizing, and using the explicit formula~\eqref{eqWigCovfh}, we
have  as our main result: 
\begin{theorem} \label{Result}
Let $\field(x,e)$ be the operator--valued distribution given by%
\footnote{We use the symbolic notation 
$a^*(\psi)=:\int d\mu(p)\, \psi(p)\circ a^*(p)$  
and \\
$a(\psi)=:\int d\mu(p)\,\overline{\psi(p)}\circ a(p)$. } 
\begin{align} \label{eqField} 
\field^\alpha(x,e) = \int_{H_m^+}\D\mu(p)\left\{  \,
e^{ip\cdot x}  \; u^\alpha(e,p) \circ a^*(p) 
+ e^{-ip\cdot x} \; \overline{u^{\bar\alpha}(e,p)} \circ a(p)\,
\right\},    
\end{align} 
with $u^\alpha$ as in~\eqref{eqIntDist}. 
Then $\field(x,e)$ is a string--localized covariant free quantum field for
$\Uirr$ in the sense of Definition~\ref{StringLocField}. 
\end{theorem}

It turns out~\cite{MSY2} that the formula works for all $\alpha \in
\CC\setminus\NN_0$, and that for a certain range of values the fields
need not be smeared in the directional variable $e$. It also works 
with  $u^\alpha$ and $u^{\bar\alpha}$ replaced by $F(p\cdot e) u^\alpha(e,p)$ 
and $\overline{F(-p\cdot e)} u^{\bar\alpha}(e,p)$, respectively, 
where $F$ is the distributional boundary value of a  suitable function which is 
analytic on the upper half plane.  
The resulting fields are all relatively ``string--local'' to each other. 
It is shown in~\cite{MSY2} that every string--localized covariant free
field, in the sense of Definition~\ref{StringLocField}, is of the
above form. 

An important open problem for our infinite spin fields is the
existence of local observables. These are operators which are localized in
bounded regions, in the sense that they commute 
with field operators localized causally disjoint from the respective region. 
In this sense, a local expression for the energy density is of 
particular interest, since it would be valuable for a discussion of
the thermodynamic properties of the KMS states~\cite{MSY} of our fields. 

We have perfomed the above construction also for massive bosons with arbitrary spin, and 
similar constructions work for fermions with half--integer spin and
for photons~\cite{MSY2}. Our photon field $A_\mu(x,e)$ is a string--localized
covariant version of the ``axial gauge'', acting on the physical
photon Hilbert space. 
The resulting fields in all these cases are strictly string--localized, but relatively local
to the corresponding standard point--localized free fields. (In fact,
they can be written as certain line integrals over the
latter~\cite{MSY2}.) 

The reason why these fields nevertheless have the
potential for applications is that they might serve as ingredients for the
construction of interacting models with string--like localization. 
Recall that the results of \cite{BGL,BuF} 
support the viewpoint that localization of charged quantum fields in 
space--like cones (the idealizations of which are our strings) 
is a natural concept, yet there is so far a lack of rigorous model 
realizations\footnote{apart from non--Lorentz covariant infra--vacua models
as in \cite{Kunhardt} and lattice models as in \cite{Mull,Lusch}.}. 
There are two reasons to believe that our free fields are good 
starting points for a 
construction of interacting
fields with strict string--localization. 
Firstly, since the obstruction to point--like localization is due to the
charge, which is already carried by the single particle states, one
should expect that already  the latter are strictly string--localized. 
That is to say, the single particle states
$E^{(1)}\field\Omega$, where $E^{(1)}$ denotes the projection onto the
single particle space and $\field$ is an interacting field, 
are string--like (but not point--like) localized in the sense of
\eqref{eqLS}. 
But then the LSZ relations imply that the 
corresponding incoming and outgoing free fields are also strictly string--localized. 
Therefore, our fields might represent the in-- and out--fields of such a
model, in contrast to the usual point--localized free fields. 
Secondly, the distributional character of our free fields is less singular
than that of the point--localized free fields, as is made
precise in~\cite{MSY2}, even more so in
the direction of the localization string. 
This fact should lead to a larger class of admissable interactions in
a perturbative construction, as compared to taking the standard
point--localized free fields as starting point.  
\appendix 
\section{Extension of the Representations to $\boldsymbol{\Poj}$} 
\label{Poj}
The proper Poincar\'e group $\Poj$ is generated by $\Po$ and any
single element $j_0$ in $\calP_+^\downarrow$. We choose 
\begin{equation}\label{eqj0}
j_0:=j_{W_0}. 
\end{equation}
(Note that $-j_0$ is in $\Po$ and hence leaves each mass shell
$H_m^+$, $m\geq 0$, invariant.) As to the irreducible
representation $\Uirr$, we choose the base point $\basepoint\in H_m^+$ so that 
\begin{equation} \label{eqjpp}
-j_0\basepoint = \basepoint.
\end{equation}
Then the section $p\mapsto A_p$ of the bundle $\Lor \to
\Lor/G_\basepoint=H_m^+$ can be chosen~\cite{MSY2} so that it transforms under 
the adjoint action of $j_0$ as 
\begin{align} \label{eqjApj} 
j_0A_pj_0=A_{-j_0p}. 
\end{align}
Let $D(j_0)$ be the anti--unitary involution from 
Lemma~\ref{UlittleJ} below. 
Then, by virtue of \eqref{eqjApj} and \eqref{eqDj}, the anti-unitary involution defined by 
\begin{align} \label{eqUj}
 \big(\Uirr(j_0)\psi\big) (p) =  D(j_0) \, \psi(-j_0p) 
\end{align}
extends $\Uirr$ to an (anti-) unitary representation of $\Poj$ within the same Hilbert space
$\Hirr=L^2(H_m^+,\D\mu)\otimes \Hlittle$. Due to irreducibility,
$\Uirr(j_0)$ is fixed up to a phase factor. 
\begin{lemma} \label{UlittleJ}
There is an anti--unitary involution $\Ulittle(j_0)$ acting on
$\Hlittle$ satisfying the representation properties%
\begin{align} \label{eqDj} 
\Ulittle(j_0)^2=1 \quad \text{ and }\quad  
\Ulittle(j_0)\Ulittle(\Lambda)\Ulittle(j_0)=\Ulittle(j_0\Lambda
j_0),\quad \Lambda\in G_\basepoint.  
\end{align}
\end{lemma}
The existence of such a representer is established in Lemma~\ref{UlittleInULor}. 
Note that the adjoint action of $j_0$ leaves $G_\basepoint$ invariant
due to \eqref{eqjpp}, hence the Lemma states that $\Ulittle$ extends
to a representation of the subgroup of $\Poj$ generated by $G_\basepoint$
and $j_0$.%
\section{Intertwiners and Localization Structure for the Principal Series Representations} 
We recall the representation of the Lorentz group presented by Bros et
al.\ in \cite{BrosMos}. 
Fix a complex number $\alpha$ with real part $-(d-2)/2$. Let $H_0^+$ denote the mantle
of the forward light cone in $\RR^{d}$ as before, and let $C^\alpha(H_0^+)$ denote the 
space of continuous $\CC$-valued functions on $H_0^+$ which are homogenous of
degree $\alpha$, i.e.
$$
C^\alpha(H_0^+):= \{\psi\in C(H_0^+):\;
\psi(tp)=t^\alpha\,\psi(p),\;t>0\}.   
$$
Consider the maps $\ULor^\alpha(\Lambda)$, $\Lambda\in\Lorj$, defined on
$C(H_0^+)$ by 
\begin{eqnarray}  \label{eqULor} 
(\ULor^\alpha(\Lambda) \psi)(p) &:=&  \psi(\Lambda^{-1}p)\,, \quad\Lambda\in\Lor \\
(\ULor^\alpha(j) \psi)(p)  &:=&  \overline{\psi(- j p)}\,,\quad j\in\calL_+^\downarrow. 
  \label{eqULorj} 
\end{eqnarray}
Clearly, $\ULor^\alpha|\Lor$ establishes a representation of $\Lor$ in
$C^\alpha(H_0^+)$, while $\ULor^\alpha|\calL_+^\downarrow$ maps
$C^\alpha(H_0^+)$ onto $C^{\bar\alpha}(H_0^+)$, and the pair
$\ULor^\alpha$, $\ULor^{\bar\alpha}$ satisfies the 
following representation  property: 
\begin{equation}\label{eqULorjl} 
\ULor^{\bar\alpha}(j_1) \ULor^{\bar\alpha}(\Lambda)\,\ULor^\alpha(j_2)
= \ULor^{{\alpha}}(j_1\Lambda j_2),
\quad \Lambda\in\Lor, j_k\in\calL_+^\downarrow.
\end{equation}
Let now $\Gamma$ be any $(d-2)$--dimensional cycle which encloses 
the origin. Then $C^\alpha(H_0^+)$ can (and will) be identified with
$C(\Gamma)$. Let $\D\nu_\Gamma$ be the restriction of the Lorentz invariant measure
$\D\nu$ on $H_0^+$ to $\Gamma$, and define a scalar product on
$C(\Gamma)$ by 
\begin{align} 
 ( \psi, \psi' ) := \int_\Gamma
\overline{\psi(p)}\,\psi'(p)\; d\nu_\Gamma(p).  
\end{align}
As Bros and Moschella point out~\cite{BrosMos}, the representation $\ULor^\alpha$ of the
Lorentz group is unitary w.r.t.\ this scalar product. The corresponding Hilbert space completion
of $C(\Gamma)$ will be denoted by $\HLor$, and the extension of $\ULor^\alpha$ to this
space will be denoted by the same symbol. It is equivalent to the irreducible principal series
representation corresponding to the value $-|\alpha|^2$ of the 
Casimir operator~\cite{BrosMos}.
\begin{lemma} \label{UlittleInULor}
i) Let $\Ulittle$ be a faithful irreducible representation of
$G_\basepoint$, and let $\Re\alpha=-(d-2)/2$. 
Then $\ULor^\alpha|G_\basepoint$ contains
$\Ulittle$, i.e.\ there is a map $T$ from a dense domain in $\HLor$
onto a dense subspace of $\Hlittle$ which
intertwines the representations $\ULor^\alpha|G_\basepoint$ and
$\Ulittle$ in the sense that 
\begin{equation} \label{eqIntULorUlittle}
 \Ulittle(\Lambda)\circ T =  T \circ \ULor^\alpha(\Lambda), \quad
\Lambda\in G_\basepoint, 
\end{equation}
holds on its domain. In the case $m>0$, 
$\Ulittle$ is a subrepresentation of $\ULor^\alpha$, while  for $m=0$, 
$\Ulittle$ occurs in a direct integral decomposition of $\ULor^\alpha$. 
$T$ also intertwines $\ULor^\alpha(j_0)$, in the sense of
\eqref{eqIntULorUlittle}, with an anti--unitary operator
$\Ulittle(j_0)$ satisfying the representation properties
\eqref{eqDj}. 
\\[1ex]
ii)  Let $R^\alpha$ be the map from (a dense domain in) 
$L^2(H_m^+)\otimes\HLor$ into $\Hirr=L^2(H_m^+)\otimes\Hlittle$
defined by 
\begin{equation} \label{eqWigCov}
\big(R^\alpha(\phi\otimes\varphi )\big)(p) :=
\phi(p)\,T\ULor^\alpha(A_p^{-1})\varphi.   
\end{equation}
Then $R^\alpha$ satisfies on its domain the intertwiner relations
\eqref{eqIntUirrUtens} and \eqref{eqIntUirrUtensJ}. 
\end{lemma}
\begin{proof}
Ad $i)$. We choose the cycle $\Gamma$ conveniently as $\Gamma
:=\Gamma_\basepoint$ defined in \eqref{eqOrbit}.%
\footnote{Note that $\Gamma_\basepoint$ is invariant under
$G_\basepoint$ and $-j_0$, which implies that the restriction of 
$\ULor^\alpha$ to the subgroup of $\Lorj$ generated by
$G_\basepoint$ and $j_0$ does not depend on $\alpha$.}  
As mentioned, the cycle $\Gamma=\Gamma_\basepoint$ 
is isometric to the sphere $S^{d-2}$ for $m>0$, and to $\RR^{d-2}$ for
$m=0$, and its isometry group coincides with $G_\basepoint$. 
Hence the action of $G_\basepoint$ on $\Gamma$ corresponds to 
the natural action of $SO(d-1)$ on $S^{d-2}$ for $m>0$, and to the
natural action of $E(d-2)$ on $\RR^{d-2}$ for $m=0$. 
It also follows that the invariant measure $\D\nu_\Gamma$ goes over into the $SO(d-2)$
invariant measure $\D\Omega$ on $S^{d-2}$ or the Lebesgue measure $\D z$ on
$\RR^{d-2}$, respectively. 
In the case $m>0$, it follows that  $\HLor$ is naturally isomorphic 
to $L^2(S^{d-2},\D\Omega)$, and $\ULor^\alpha|G_\basepoint$ acts as 
the  push--forward representation. 
As is well-known, this representation decomposes into the direct sum of
all irreducible representations $\Ulittle_{(s)}$ of $SO(d-1)$. 
(In $d=4$, $s$ runs through  $\NN_0$ and the irreducible subspaces 
are spanned by the spherical harmonics $Y_{s,m}$, and in $d=3$,  $s$ runs through $\ZZ$ 
and the irreducible subspaces are spanned by $\theta\mapsto \exp(\I s\theta)$.)  
Hence, for each $s$ there is a partial isometry $T=T_{(s)}$ with the
claimed property \eqref{eqIntULorUlittle}. 
Further, under the mentioned equivalence $\Gamma\cong S^{d-2}$ the
representer of $j_0$ acts as $(\ULor^\alpha(j_0)\varphi)(n)=
\overline{\varphi(I_0n)}$, where $I_0$ corresponds to $-j_0$ and is
hence in $O(d-1)$. Since the spherical harmonics 
$\{Y_{s,m},m=-s,\ldots s\}$ for given $s\in\NN$ are
invariant under $O(3)$ as well as under complex conjugation, 
it follows that $\ULor^\alpha(j_0)$ leaves each irreducible subrepresentation
invariant in d=4. This implies that $\ULor^\alpha(j_0)$ is intertwined by $T$
with an (anti-unitary) operator $\Ulittle(j_0)$ satisfying
\eqref{eqDj}, as claimed. 
In d=3, the same 
conclusion follows from the facts that $I_0$ is an orientation
reversing isometry of the circle, hence $SO(2)$-conjugate to
$\theta\mapsto -\theta$, and that $\overline{\exp(\I
s(-\theta))}=\exp(\I s\theta)$. 

Similarly, in the case $m=0$,  $\HLor$ is naturally isomorphic to
$L^2(\RR^{d-2},\D z)$, and $\ULor^\alpha|G_\basepoint$ acts as  
the push--forward representation. 
Via Fourier transformation, this representation decomposes into a direct integral of
irreducible representations $\Ulittle_{(\kappa)}$, where $\kappa$ runs
through $\RR$ for $d=3$, and through $\RR^+$ for $d=4$. 
Thus there is a densely defined intertwiner $T$ 
satisfying  \eqref{eqIntULorUlittle}  on its
domain: $T\varphi$ is the restriction of the Fourier transform of
$\varphi$ to the circle with radius $\kappa$ for $d=4$, respectively
its value at $\kappa$ for $d=3$.    
Further, under the mentioned equivalence $\Gamma\cong \RR^{d-2}$ the
representer of $j_0$ acts as $(\ULor^\alpha(j_0)\varphi)(z)=
\overline{\varphi(I_0z)}$, where $I_0$ corresponds to $-j_0$.  
With our explicit formula \eqref{eqGammaR}, $I_0$ coincides  with the
reflection $z\mapsto -z$.%
\footnote{Using another isometric diffeomorphism yields the same $I_0$
up to conjugation with a euclidean transformation, leading to the same
conclusion.} 
The identity $T\ULor^\alpha(j_0)\varphi=\overline{T\varphi}$ then implies
that $\ULor^\alpha(j_0)$ leaves the kernel of $T$ invariant. Hence   
$$
D(j_0) T \varphi := T\,\ULor^\alpha(j_0) \varphi,
$$ 
defines an anti--unitary operator $D(j_0)$ on the image of $T$, which also has
the representation property~\eqref{eqDj}, as claimed. 

Ad $ii)$. The intertwiner relations \eqref{eqIntUirrUtens} and
\eqref{eqIntUirrUtensJ}  follow from part $i)$, \eqref{eqjApj} and \eqref{eqULorjl}. 
\smartqed \qed
\end{proof}
We now discuss the map $F^\alpha$ defined by Bros et al.\ \cite{BrosMos}, 
which we used in Example~\ref{LSULor}. It is the Fourier-Helgason type transformation  given by 
\begin{align} 
F^\alpha & : \calD(\Spd^{d-1})\rightarrow C^\alpha(H_0^+)\subset \HLor ,\\
(F^\alpha h)(p)&:= \E^{-\I\pi\alpha/2}\;
\int_{\Spd^{d-1}} \D\sigma(e) \,h(e)\,(e\cdot p)^\alpha.  \label{eqFH}
\end{align}
Here,  $e\cdot p$ denotes the scalar product in $d$-dimensional 
Min\-kow\-ski space, of which $\Spd^{d-1}$ and $H_0^+$ are considered
submanifolds. The power $t^\alpha$ is defined via the branch of the  
logarithm on $\RR\setminus \RR^-_0$ with $\ln 1=0$, and as 
$\lim_{\varepsilon\rightarrow 0+}(t+i\varepsilon)^{\alpha}$ for $t<0$. 
Further, $\D\sigma$ denotes the Lorentz invariant measure on
$\Spd^{d-1}$. 
In our context, the upshot of this transformation is the following. 
\begin{lemma}[Bros et al.] \label{FHProp'}
The map $h\mapsto F^\alpha h$ establishes a ``localized covariant wave function''
on $\Spd^{d-1}$ in the sense of the properties $0)\ldots ii)$ listed in Example~\ref{LSULor}. 
\end{lemma}
\begin{proof}
The transformation $F^\alpha$ has been taken over from \cite{BrosMos}
in such a way that $F^\alpha h=\phi(h)\Omega$, where $\phi(\cdot)$ is
the free field of~\cite{BrosMos},  cf.~\cite[eq.~(4.30)]{BrosMos}. 
In this context, property $0)$ of our Example~\ref{LSULor} is the Reeh--Schlieder property, 
Proposition~5.4 of \cite{BrosMos}. The covariance
property $i)$ corresponds to the covariance of the field $\phi(\cdot)$
(and also follows directly from the definitions). 
Finally, the geometrical KMS condition~\cite[Prop.~2.3]{BrosMos}  
enjoyed by the two-point function of $\phi(\cdot)$ implies that
$F^\alpha h$ is in the domain of the Tomita operator  $\exp(-\pi
K^\alpha_{{W}})$. The antipodal condition~\cite[Prop.~2.4]{BrosMos}
then shows that this operator acts on $F^\alpha h$ as in \eqref{eqFHModOp}. 
This proves $ii)$.  
\smartqed\qed
\end{proof}
%
We finally mention a standard result,  which we have used occasionally
in the context of our modular operators. 
\begin{lemma} \label{ModOp} 
Let $U_t$ be a continuous unitary one-paramter group, with generator $K$. Then
$\psi$ is in the domain of $\exp(-\pi K)$ if, and only if, the
vector-valued map 
$$ 
t\mapsto U_t\psi 
$$
is analytic in the strip $\RR+i(0,\pi)$ and weakly continuous on the
closure of that strip. In this case, $\exp(-\pi K)\psi$ coincides with
the analytic continuation of $U_t\psi$ into $t=i\pi$. 
\end{lemma}
%
%



\printindex
\end{document}